\documentclass[aip, pof, amsmath, amssymb, preprint]{revtex4-2}
\usepackage[english]{babel}
\usepackage[utf8]{inputenc}
\usepackage{bm}
\usepackage{amsthm}
\usepackage{mathtools}
\usepackage{physics}
\usepackage[dvipsnames]{xcolor}
\usepackage[colorlinks]{hyperref}
\hypersetup{citecolor=Blue, urlcolor=Blue, linkcolor=Blue, filecolor=black}
\usepackage{graphicx}
\usepackage[left=23mm,right=13mm,top=35mm,columnsep=15pt]{geometry} 
\usepackage{adjustbox}
\usepackage{placeins}
\usepackage[T1]{fontenc}
\usepackage{lipsum}
\usepackage{csquotes}
\usepackage{upgreek}
\usepackage{multirow}
\usepackage[separate-uncertainty=true]{siunitx}
\usepackage{natbib}
\usepackage{array}
\AtBeginDocument{}
\makeatletter
\let\oldtheequation\theequation
\renewcommand\tagform@[1]{\maketag@@@{\ignorespaces#1\unskip\@@italiccorr}}
\renewcommand\theequation{(\oldtheequation)}
\makeatother
\AtBeginDocument{}

\usepackage{comment}

\begin{document}
\title{Damage and recovery of flagella in soil bacteria exposed to shear within long microchannels}

\author{Juan Pablo Carrillo-Mora}
    \altaffiliation[Current address: ]{Departament de Física de la Matèria Condensada and UBICS University of Barcelona Institute of Complex Systems, Universitat de Barcelona, Barcelona, Spain}
    \email{jpcarrillo-mora@ub.edu}
    \affiliation{Departamento de Física - Facultad de Ciencias Físicas y Matemáticas, Universidad de Chile, Santiago, Chile}
\author{Moniellen Pires Monteiro}
    \affiliation{Departamento de Física - Facultad de Ciencias Físicas y Matemáticas, Universidad de Chile, Santiago, Chile}
\author{Aníbal R. Lodeiro}
    \affiliation{Instituto de Biotecnología y Biología Molecular - Facultad de Ciencias Exactas and CCT-La Plata CONICET, Universidad Nacional de La Plata, La Plata, Argentina}
    \affiliation{Laboratorio de genética - Facultad de Ciencias Agrarias y Forestales, Universidad Nacional de La Plata, La Plata, Argentina}
\author{V. I. Marconi}
    \affiliation{Facultad de Matemática, Astronomía, Física y Computación, Universidad Nacional de Córdoba, Córdoba, Argentina}
    \affiliation{IFEG-CONICET, X5000HUA, Córdoba, Argentina}
\author{María Luisa Cordero}
    \email{mcordero@ing.uchile.cl}
    \affiliation{Departamento de Física - Facultad de Ciencias Físicas y Matemáticas, Universidad de Chile, Santiago, Chile}

\date{\today}

\begin{abstract}
The swimming motility of bacteria is driven by the action of bacterial flagellar motors, whose outermost structure is a long and thin helicoidal filament. When rotated, the fluid medium exerts an anisotropic viscous drag on the flagellar filaments, ultimately leading to bacterial propulsion. The flagellar filaments are protein-based flexible structures that can break due to interactions with fluid flows. Here, we study the evolution of flagellar filaments in the soil bacterium \textit{Bradyrhizobium diazoefficiens} after being exposed to shear flows created in long microchannels, for shear rates between \SI{1}{\per\second} and $10^5~\si{\per\second}$, and for durations between tens of milliseconds and minutes. We demonstrate that the average swimming speed and fraction of swimming cells decrease after exposition to shear, but both parameters can recover, at least partially, with time. These observations support the hypothesis that shear flows cut flagellar filaments but that reversibly damaged bacterial flagellar motors can be restored thanks to filament regeneration. By fitting our observations with phenomenological expressions, we obtain the individual growth rates of the two different flagellar filaments that \textit{B. diazoefficiens} possesses, showing that the lateral filaments have a recovery time of about \SI{40}{\minute} while the subpolar one requires more than \SI{4.5}{\hour} to regrow. Our work demonstrates that simple monitoring of bacterial motility after exposition to shear can be used to characterize the process of flagellar filament breakup and growth, a phenomenon widely present in bacteria swimming in porous soil and exposed to shear flows due to rainfall and watering systems.
\end{abstract}

\maketitle

\section{Introduction}

The bacterial flagellar motor (BFM) is a complex protein-based structure responsible for swimming motility. Despite the large diversity among different species, the basic structure is highly conserved and consists of a basal body that includes a stator and a rotor, and a long helicoidal filament, joined to the basal body through a flexible hook~\cite{Minamino2015}. The formation of the BFM proceeds from the inside-out in a wonderfully orchestrated process~\cite{Macnab2003}. Once the basal body and hook are formed, the filament grows through an injection-diffusion mechanism~\cite{56renault2017, 57chen2017}, in which filament sub-units are injected through the export apparatus and diffuse in a single line through the hollow structure of the filament until they reach the free tip, where they immediately crystallize, progressively extending the filament. After its assembly, the motive force of the BFM is the crossing of protons or ions across the membrane, causing shape changes in the stator and the consequent rotation of the rotor. Such movement is transmitted to the hook and the helical filament. Following this rotation, the filament experiences an anisotropic viscous drag from the outer liquid medium, resulting in the propulsion of the cell in the surrounding viscous medium~\cite{Lauga2009}. The viscous drag on the filament is countered by an equal (in magnitude) but opposite (in direction) drag on the bacterial body, thus ensuring that the whole bacterium is force-free, as required by the extremely small Reynolds numbers associated with microswimmer flows~\cite{Purcell1977}.

In contrast to the rest of the components of the motor, which have a precisely defined number and proportion of the conformational proteins, the filament grows continuously and can be formed by a variable number of filament sub-units. The resulting swimming speed of a bacterium is then proportional to the rotational frequency of the flagella, to their number, and to the length of their filaments, at least as long as their length is below an optimal value~\cite{53higdon1979, 54higdon1979, Spagnolie2010, 52qin2012, Liu2014, Nguyen2018, lisevich2024}.

Long flagellar filaments can be easily broken due to interactions with fluid flow~\cite{Turner2012, Paradis2017}. For example, soil bacteria are constantly exposed to external flows produced by rainfall or irrigation systems, which can cut their flagella. In such cases, continuous growth of the flagellar filament enables a recovery mechanism to ensure sustained motility. To study the flagellar damage and recovery, in this work, we conducted experiments where bacteria were exposed to a controlled shear flow inside long microchannels, and its effect on motility, specifically on the fraction of swimming bacteria and their swimming speed, was observed.

Our experiments used a natural soil bacterium, the nitrogen-fixing symbiont of the soybean, \textit{Bradyrhizobium diazoefficiens}. Besides its relevance in sustainable agriculture as a biofertilizer~\cite{Catroux2001}, \textit{B. diazoefficiens} is an interesting model microswimmer as it possesses two different flagellar systems: the subpolar system, consisting of a single thick flagellum, always expressed, located near one pole of the bacterium body, and the lateral one, composed of several thin and long flagella peritrichously located on the bacterium body that can be produced depending on the available carbon source and viscosity of the medium~\cite{19quelas2016, 20garrido2019}. Although evolutionarily distinct, both flagellar systems crosstalk, with the subpolar flagellum acting as a mechanosensor that regulates the expression of the lateral flagella~\cite{18mengucci2020}. The presence of both flagellar systems appears to confer \textit{B. diazoefficiens} an advantage for swimming motility, at least far from solid boundaries, as the average swimming speed is reduced in mutants devoid of one of the two flagellar systems and especially in the absence of the subpolar flagellum~\cite{Althabegoiti2011, 19quelas2016, gutierrezMaster2023, MonteiroPreprint}. This evidence suggests that both types of flagella can have different responses to environmental conditions~\cite{cortezMaster2014, montagnaMaster2018} and regeneration rates. To test this hypothesis, we expose two strains of \textit{B. diazoefficiens} to shear, the wild type (WT) expressing both flagellar systems, and a mutant strain ($\Delta$\textit{lafA}) that only has the subpolar flagellar system, and monitor their behavior after various shearing conditions (Fig.~\ref{fig: microchannels}(a)). In this way, we decouple the recovery of both flagellar systems. Overall, our study gives insight into the damage that fluid shear can cause to the BFM and enables a simple yet powerful means to monitor flagellar growth from the main motility parameters of the bacterial population.

\section{Methods}

\subsection{Culture protocols}

Two different strains of \textit{B. diazoefficiens} USDA 110, obtained from the United States Department of Agriculture, Beltsville, were used: the WT has both flagellar systems, while the $\Delta$\textit{lafA} is a mutant with a genomic deletion that can only express the subpolar flagellum~\cite{Althabegoiti2011} (Fig.~\ref{fig: microchannels}(a)). For routine use, bacterial stocks were maintained at \SI{4}{\celsius} in solid yeast extract mannitol-agar medium (YEM agarized at \SI{1.5}{\percent})~\cite{58dardis2021}, which were renewed every three months. For the WT strain, Chloramphenicol (\SI{20}{\milli\gram/\litre}) was used for the semi-solid medium. For the experiments, cultures in liquid growth medium HMY-arabinose~\cite{18mengucci2020} were initiated from a single colony in the agar plates and grown at a temperature of \SI{28}{\celsius} and 180 rpm of agitation to late log phase, until reaching an optical density at \SI{600}{\nano\meter} wavelength $\text{OD}_{600} = 3.0 \pm 0.1$~\cite{39beal2020}. Bacteria were then diluted in \SI{10}{\milli\litre} of the same medium, setting the initial $\text{OD}_{600}$ to 0.1. This reculture was grown at the same temperature and agitation until it reached $\text{OD}_{600} = 1.0 \pm 0.1$. Finally, a 1:5 dilution of the reculture in HM-salts supplemented with \SI{5}{\percent} L-arabinose and PVP-40 (Polyvinylpyrrolidone-40, Sigma Aldrich) at \SI{0.05}{\percent} w/v (HM-Ara-PVP) was kept at rest without agitation at \SI{28}{\celsius} for \SI{6}{\hour} before inoculation into the microfluidic device for optimal motility. By using PVP-40 in the minimal medium solution, adhesion of the bacteria to the walls was prevented, and self-agglutination was reduced~\cite{51jiang2023}. Despite observations indicating that lateral flagella are adapted for swimming in viscous media~\cite{18mengucci2020}, we expect that the small concentration of PVP-40 used here affected negligibly the bacterial behavior.

\subsection{Microfluidic devices and shearing conditions}

\begin{figure*}
    \centering
    \includegraphics[width=15.5cm]{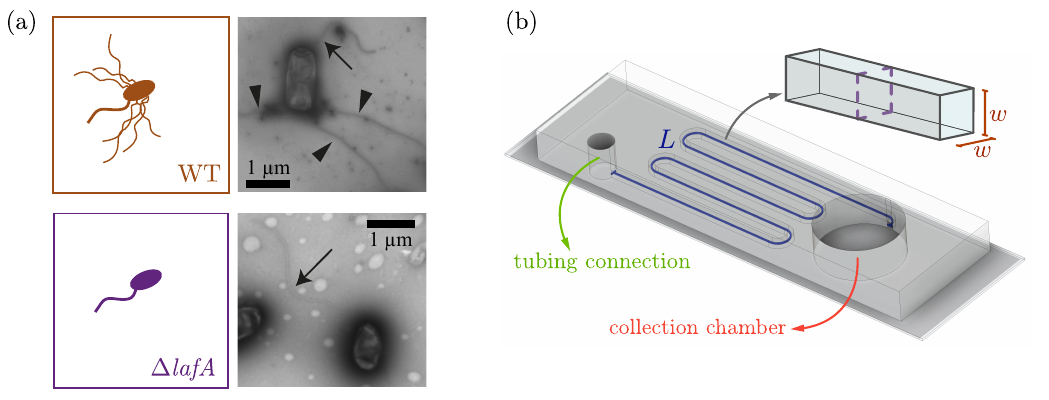}
    \caption{(a) The two strains of \textit{B. diazoefficiens} used in this study. The wild type possesses two types of flagella: the subpolar (marked with an arrow in the TEM images) and the lateral (marked with an arrowhead in the TEM image). The mutant strain $\Delta$\textit{lafA} only expresses the subpolar one. (b) Schematic of the microfluidic devices used in the experiments. The devices consist of a square cross-section microchannel of length $L$ and height/width $w = \SI{15}{\micro\meter}$. One end of the microchannel is connected to a tubing through which the bacterial suspension is injected, and the other is connected to a collection chamber where the sheared suspension is recovered.}
    \label{fig: microchannels}
\end{figure*}

Microfluidic devices were fabricated using maskless optical and soft lithography~\cite{menon2005, qin2010}. They consist of thin microchannels of square cross section, $w = \SI{15}{\micro\meter}$ in width/height, and a controlled length $L$ that ranges between \SI{4.5}{\milli\meter} and \SI{2.25}{\meter}. For lengths $L$ longer than \SI{45}{\milli\meter}, the microchannels are bent as a serpentine to fit in a microscope slide. A schema of a serpentine microchannel geometry is shown in Fig.~\ref{fig: microchannels}(b). 

In the first set of experiments, bacteria were exposed to flows over a constant distance in a straight microchannel of length $L = \SI{45}{\milli\meter}$. The injection flow rate $Q$ was varied to regulate the mean shear rate~\cite{30figueroa2015} $\dot{\gamma} = 2Q/w^3$. Since the flow rate was increased to increase $\dot{\gamma}$, the time for which the microorganisms were sheared varied in each case, decreasing for increasing $\dot \gamma$.

In the second set of experiments, the samples were sheared in microchannels of different lengths, designed for each target average shear rate to maintain a constant shear time, as explained next. The shearing time depends on the flow as $t_s = w^2 L/Q$, which in terms of the mean shear rate is $t_s = 2 L/(\dot{\gamma} w)$. Thus, to maintain a constant shear time $t_s$, the microchannel length had to be proportional to the desired $\dot \gamma$. The shearing time was fixed at $t_s = \SI{60}{\second}$, and the microchannel lengths were adjusted accordingly.

A summary of experimental condition, i.e., mean shear rates $\dot{\gamma}$, channel lengths $L$, injection flow rates $Q$, and shearing times $t_s$, are shown in Tab.~\ref{tab: constant shearing time}. Note that the maximum shear rate we could attain in the experiments with constant shearing time was lower than in the experiments with constant shearing distance. The limitation was technical: the longest channel was over \SI{2}{\meter} in length, requiring a very high pressure for its operation. For longer channel lengths, the tubing used for injection did not withstand such high operating pressures and unplugged from the microchannels.

\begin{table*}[t]
	\begin{tabular}{p{2cm}p{2cm}p{2cm}p{2cm}p{2cm}p{2cm}p{2cm}}
	& \multicolumn{3}{c}{Constant shearing distance} & \multicolumn{3}{c}{Constant shearing time} \\
	\hline
	$\dot \gamma$ (\si{\per\second}) & $L$ (\si{\milli\meter}) & $Q$ (\si{\nano\litre/\second}) & $t_s$ (\si{\second}) & $L$ (\si{\milli\meter}) & $Q$ (nL/s)  & $t_s$ (s)  \\
	\hline
	$1 \times 10^1$    & 45 & 0.0168 & 600  &  4.5   & 0.0168 & 60 \\
	$1 \times 10^2$    & 45 &  0.168  &  60   &   45   &  0.168  & 60 \\
	$1 \times 10^3$    & 45 &   1.68   &   6    &  450  &   1.68   & 60 \\
	$0.5 \times 10^4$ &  --  &     --     &   --    & 2250 &   8.44   & 60 \\
	$1 \times 10^4$    & 45 &   16.8   &  0.6  &    --    &      --    & -- \\
	$1 \times 10^5$    & 45 &   168    & 0.06 &    --    &      --    & -- 
	\end{tabular}
    \caption{Summary of experimental conditions.}
    \label{tab: constant shearing time}
\end{table*}

For the control $\dot{\gamma}_0 = 0$, the motility was measured directly in a cylindrical microfluidic cavity without exposing bacteria to flow.

\subsection{Inoculation and data acquisition}

To prepare the microfluidic device for the experiments, one end of the tubing was connected to the microchannel, while HM-Ara-PVP was injected through the other end using a glass syringe and syringe pump (neMESYS Base 120 and low-pressure module V2, Cetoni). Once the microchannel was filled, ensuring no air bubbles were left inside, the injection was stopped, and the collection chamber was emptied so that only the microchannel was flooded. Then, the syringe was disconnected and replaced by another one containing the bacterial suspension used in the experiment. Finally, the bacterial suspension was injected through the tubing with a constant flow rate, going through the microchannel until it arrived at the collection chamber (Fig.~\ref{fig: microchannels}(b)). Once filled, the sheared suspension was transferred to a large cylindrical cavity to avoid flow, where bacterial swimming was observed.

The bacteria in the cylindrical microfluidic cavity were observed with a Nikon Eclipse TS100 microscope in a bright field, a 40$\times$/0.6 NA objective, and an Andor Zyla sCMOS camera. Videos were recorded at 50 fps (frames per second), with an extension of 1000 frames and a resolution of 1024 $\times$ 1024 pixels$^2$. Calibration for the optical system indicates a ratio of 6.24 pixel/\si{\micro\meter}, this implies that the imaged regions have an area of 164 $\times$ 164 \si{\micro\meter^2}.

Measurements for all shear rates were performed with three biological replicas of each strain.

\subsection{Fraction of active bacteria and swimming speed}

\begin{figure}[h]
\centering
\includegraphics{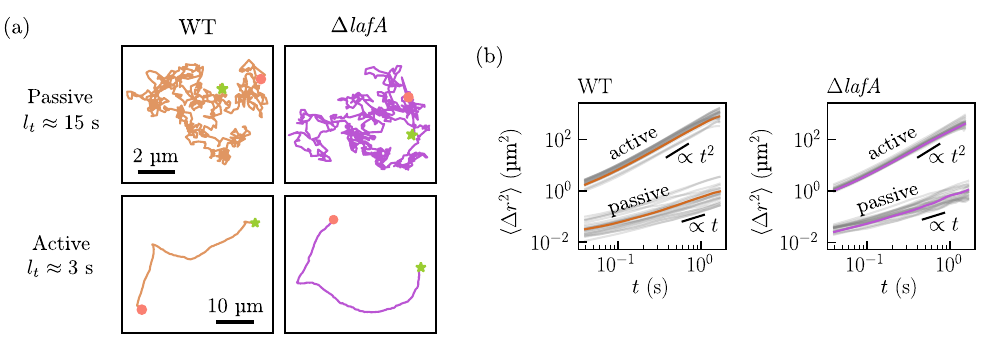}
\caption{(a) Examples trajectories for passive (top) and active (bottom) cells of WT (left) and $\Delta$\textit{lafA} (right). The green stars and orange dots indicate the initial and end points of the respective trajectory. The scale bar and temporal length $l_t$ of the trajectories are the same for each row. (b) MSD for active and passive bacteria. Thin gray lines show several individual example curves. The thick color curves (brown for WT and violet for $\Delta$\textit{lafA}) correspond to the MSD of the example trajectories shown in (a).}
\label{fig: active vs passive}
\end{figure}

The videos obtained from the experiments were processed with Fiji (ImageJ) software~\cite{43schindelin2012} to enhance their brightness and contrast, and then analyzed with our open source software \textit{Biotracker}~\cite{44sanchez2016, reyesMaster2017, biotracker2017} to obtain the trajectories of the bacteria. Trajectories revealed two markedly different bacterial motility patterns. The ``active'' bacteria self-propelled and swam for considerable distances in little time intervals, whereas the ``passive'' bacteria showed erratic and fluctuating movement reminiscent of Brownian motion (see Fig.~\ref{fig: active vs passive}(a)). This is consistent with previously reported swimming behavior of \textit{B. diazoefficiens}~\cite{19quelas2016, cortezMaster2014, montagnaMaster2018} and also common in other bacterial species~\cite{Mino2011}.

To distinguish between active and passive bacteria, we computed the mean-squared displacement (MSD) for each trajectory as
\begin{equation}
\langle \Delta r^2 (t) \rangle = \left\langle (x(t_0 + t) - x(t_0))^2 + (y(t_0 + t) - y(t_0))^2 \right\rangle,
\end{equation}
where $\langle \cdot \rangle$ represents an average over all possible initial times in the trajectory, $t_0$. The MSDs of active and passive bacteria were segregated and grew quadratically and linearly with time lag $t$, respectively, as shown in Fig.~\ref{fig: active vs passive}(b). This allowed us to establish a robust criterion to distinguish between active and passive bacteria. Accordingly, the fraction of active bacteria was defined as 
\begin{equation}
\varphi = N_A/(N_A + N_P),
\end{equation}
where $N_A$ and $N_P$ were the number of active and passive bacteria in the suspension, respectively.

The behavior of the MSD curves was consistent with the theoretical MSD of a self-propelled particle subjected to thermal noise in two dimensions~\cite{Howse2007, Martens2012}
\begin{equation}
\langle \Delta r^2 (t) \rangle = \left( 4 D_T + 2 v^2 \tau \right)t + 2 v^2 \tau^2 \left( e^{-t/\tau} - 1 \right),
\label{eq:ActiveMSD}
\end{equation}
where $D_T$ is the diffusion coefficient associated with thermal fluctuations in the liquid medium, $v$ is the self-propulsion speed, and $\tau$ is the characteristic reorientation time, which combines the effects of thermal rotational diffusion and bacterial reorientations. The agreement between the experimental MSD curves and Eq.~\eqref{eq:ActiveMSD} is satisfactory for time lag $t \gtrsim \SI{0.5}{\second}$, below which detection errors in the position of bacteria decrease the power-law exponent~\cite{Martin2002}. For passive particles, $v = 0$ and $\langle \Delta r^2 (t) \rangle = 4 D_T t$, thus allowing us to obtain a diffusion coefficient $D_T = \SI{0.24 \pm 0.02}{\micro\meter^2/\second}$, consistent with the expected diffusion coefficient of a Brownian particle of diameter \SI{1}{\micro\meter} (the error corresponds to the standard deviation of all diffusion coefficients measured for passive bacteria). For active particles, provided that $t \ll \tau$, one obtains
\begin{equation}
\langle \Delta r^2 (t) \rangle \approx 4 D_T t + v^2 t^2.
\end{equation}
Thus, by fitting each of the MSD curves of active bacteria (gray curves in Fig. ~\ref{fig: active vs passive}(b)), their individual swimming speed $v$ was obtained.

\section{Results}

\subsection{Motility decay}

\begin{figure}[ht!]
    \centerline{\includegraphics{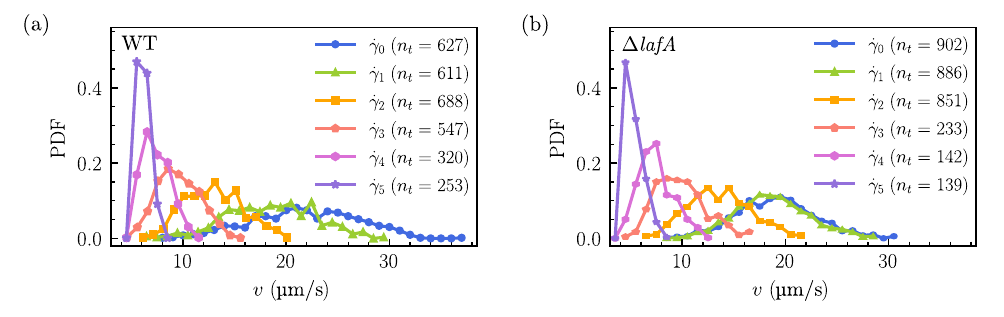}}
    \caption{Probability density functions for the swimming speed of strains (a) WT and (b) $\Delta$\textit{lafA} after being exposed to flows with mean shear rates $\dot{\gamma}_i = 10^i~\si{\per\second}$ (where $i = 1, 2, \dots, 5$) in the experiments with constant length. The control $\dot{\gamma}_0$ corresponds to suspensions not exposed to flows. The number of trajectories ($n_t$) used to calculate the PDFs is indicated in each case.
    }
    \label{fig: vel dist shear}
\end{figure}

The probability density function (PDF) of the swimming speed for active bacteria sheared for constant length with varying mean shear rates are shown in Fig.~\ref{fig: vel dist shear}. A qualitatively similar result was obtained for constant shearing time (not shown). For both strains, both the mean and standard deviation of the swimming speed distributions decreased as the mean shear rate increased. That is, as the shear stresses to which the bacteria were subjected increased, they swam slower, and the variability between the swimming speeds of the bacteria in the suspension also decreased. Since only the swimming speed of active bacteria was considered for constructing the distributions shown in Fig.~\ref{fig: vel dist shear}, the decrease in the mean swimming speed was not due to an increase in the number of passive bacteria in the suspension. Instead, the active bacteria were self-propelled at a lower speed.

\begin{figure}[ht!]
    \centerline{\includegraphics{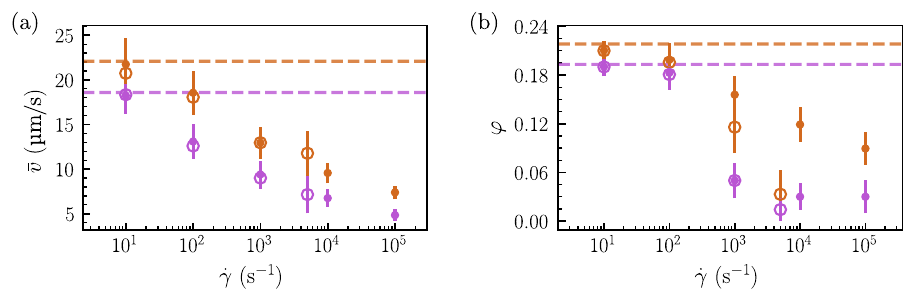}}
    \caption{Mean swimming speed (a) and active fraction (b) of WT (brown) and $\Delta$\textit{lafA} (violet) bacteria as a function of the mean shear rate to which they were exposed. Solid and open symbols correspond to experiments performed at constant shearing length and constant shearing time, respectively. The data points represent the averages of the measured values, and their error bars correspond to the standard error. The horizontal dashed lines correspond to the values measured in the control, non-sheared suspensions.
    }
    \label{fig: v_m and active fraction shear}
\end{figure}

The averages of the swimming speed distributions, $\bar v$, are shown in Fig.~\ref{fig: v_m and active fraction shear}(a) for both constant shearing length (solid symbols) and constant shearing times (open symbols). Up to the highest mean shear rate studied in the experiments with constant shearing time, no systematic differences were apparent in the drop of mean swimming speed between the experiments with constant shearing length versus constant shearing time.

On the other hand, Fig.~\ref{fig: v_m and active fraction shear}(b) shows the active fraction, $\varphi$, of each strain as a function of the mean shear rate of the flow to which they were exposed. It can be seen that for both strains, there was a slow decay of $\varphi$ up to shear rates \SI{1e2}{\per\second}, and once this critical shear rate was exceeded, the active fraction decayed rapidly for both strains. The decrease in the active fraction is also evident from the decreasing number of active bacteria available for constructing the PDFs of Fig.~\ref{fig: vel dist shear} in the recorded videos. A notorious difference between strains was observed when comparing the experiments performed with constant shearing distance versus constant shearing time. The decrease in the active fraction of WT was much more pronounced when the shearing was applied for a constant time in comparison with a constant shearing length, whereas the active fraction of the $\Delta$\textit{lafA} decayed similarly, independent of the shearing conditions.

In these experiments, it was only possible to visualize the body of the bacteria moving while they swam, i.e., it was not possible to see their flagella. However, based on the results reported in other works~\cite{52qin2012, 54higdon1979, 55magariyama1995, Turner2012, lisevich2024}, a plausible hypothesis can be put forward to explain the decrease in the mean swimming speed and active fraction observed in both strains with increasing $\dot{\gamma}$ (Fig.~\ref{fig: v_m and active fraction shear}). Considering that the swimming speed is proportional to the length and number of flagellar filaments~\cite{53higdon1979, 54higdon1979, Nguyen2018}, at least part of the BFMs may have been damaged, either reversibly by a partial or total cut of the flagellar filaments or irreversibly because of a compromise of the more internal flagellar structures. Since the WT strain possesses two flagellar systems, a complete cut of filaments and/or irreversible flagellar damage on some (but not all) of the BFMs of a cell would still enable it to swim, albeit with a lower speed. Conversely, a WT cell with all of its flagellar filaments completely cut and/or BFMs irreversible damaged would become a passive bacterium. The $\Delta$\textit{lafA} strain, on the other hand, has a single flagellum. Thus, a partial filament breakup would decrease its swimming speed, while a complete filament cut or irreversible damage on the BFM of a cell would render it passive. The fundamental difference between a completely cut filament and an irreversibly damaged BFM is the possibility of recovery of the broken filament. Thus, a bacterium that became passive following the complete mutilation of all its flagellar filaments but whose internal flagellar structures remained intact would eventually recover motility and become an active bacterium again.

Based on this hypothesis, we postulate that higher shear rates may have cut more flagellar filaments and/or irreversibly damaged more BFMs, causing a larger drop in the bacteria's swimming speed and active fraction. In the case of the WT strain, both effects could explain the decrease in swimming speed, while in the case of the $\Delta$\textit{lafA} mutant, the reduction of swimming speed shown in Fig.~\ref{fig: v_m and active fraction shear}(a) would reflect only the partial breakup of its flagellar filament. The similarity of the velocity decrease in the experiments with constant shearing distance and constant shearing time suggests that, for both types of flagella, the average length and number distribution of remaining filaments does not depend on the time and distance of exposure to the flow, but only on its shear rate.

On the other hand, the drop in the fraction of active bacteria in WT and $\Delta$\textit{lafA} suspensions as a function of shear rates (Fig.~\ref{fig: v_m and active fraction shear}(b)) could be understood as an increase in bacteria whose flagella were completely damage, either reversibly or irreversibly. In general, the drop in active fraction for the WT strain was less pronounced than for the $\Delta$\textit{lafA} mutant with increasing shear rates, which suggests that the presence of lateral flagella helped protect bacteria from completely losing their motility when exposed to external flows, either because their lateral flagella are more resistant to shear stresses or because they modify the coupling with the flow, reducing the probability of flagellar damage. This is consistent with the observation that the decrease in active fraction for the $\Delta$\textit{lafA} was comparable in the experiments with constant shearing distance and constant shearing time, suggesting that the mere exposition to shear can produce significant damage to the subpolar flagellum if it is not protected by the lateral flagella, independent of the duration of the shearing. In the case of the WT strain, on the other hand, the decrease in the active fraction was significantly higher in the experiments with constant shearing time, suggesting increased damage to lateral flagella if the shearing is sustained for a considerable amount of time, in this case for \SI{60}{\second}.

\subsection{Motility recovery}

Because the broken flagellar filaments of bacteria can regenerate~\cite{56renault2017, 57chen2017, Paradis2017}, the hypothesis that the shear on bacteria exposed to an external flow can cut their flagella, decreasing their swimming speed, can be further explored by studying the evolution of the motility of a sheared suspension. Then, if indeed the flagellar filaments of bacteria were cut by their exposure to the flow, causing the drop of $\bar v$ and $\varphi$, a recovery of motility should be observed as the filaments regenerate, which could be quantified by measuring the time evolution of these parameters after being sheared. 

\begin{figure}[ht!]
    \centering
    \includegraphics{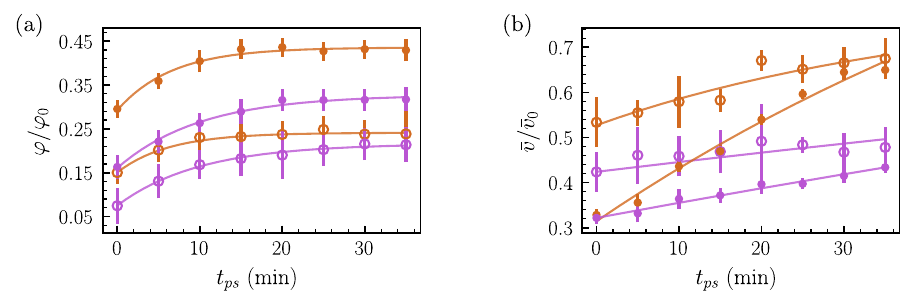}
    \caption{Recovery of post-shear motility of suspensions of WT (brown) and $\Delta$\textit{lafA} (violet) strains exposed to flows with mean shear rates $\dot{\gamma} = \SI{1e5}{\per\second}$ (constant shearing distance, solid symbols) and $\dot \gamma = \SI{0.5e4}{\per\second}$ (constant shearing time, open symbols). (a) The active fraction of bacteria as a function of post-shear time $t_\text{ps}$, normalized by the active fraction of non-sheared bacteria, $\varphi_0$. (b) Mean swimming speed of bacteria as a function of post-shear time $t_\text{ps}$, normalized by the mean swimming speed of non-sheared bacteria, $\bar v_0$. The symbols correspond to the average of the measured values, and their error bars correspond to the standard error. The continuous lines shown in the figures correspond to the curves of the expression fitted to the experimental data, Eq.~\eqref{eq: model phi_A recovery} in (a) and Eq.~\eqref{eq: model to fit WT v2} in (b).}
    \label{fig: recovery mot}
\end{figure}

Following the above, the time evolution of the motility parameters $\bar v$ and $\varphi$ was studied in the suspensions that were exposed to the flows with the highest mean shear rates ($\dot{\gamma} = \SI{1e5}{\per\second}$ in the experiments with constant shearing distance and $\dot{\gamma} = \SI{0.5e4}{\per\second}$ in the experiments with constant shearing time), since they corresponded to the cases with the highest loss of motility. For both strains, partial recovery of the fraction of active bacteria in the suspensions and their mean swimming speed was observed, as shown in Fig.~\ref{fig: recovery mot}.

Measurements of the active fraction of these suspensions as a function of post-shearing time, $t_\text{ps}$, exhibited partial recovery, which saturated at lower values than the control (see Fig.~\ref{fig: recovery mot}(a)). The saturation values were consistently lower for the experiments with constant shearing times, implying that a larger number of cells had all of their flagella irreversibly damaged when the shearing was sustained for a longer time. Similarly, the mean swimming speed of the sheared suspensions of both strains was recovered with post-shear time (see Fig.~\ref{fig: recovery mot}(b)). In this case, no saturation of the mean swimming speed was observed for the duration of the experiment (\SI{35}{\minute}). However, the curves seemed to converge to similar saturation values for each strain, irrespective of the shearing conditions. This suggests that bacteria that retain functional flagella after the shearing would eventually recover similar motility levels, regardless of the shearing conditions.

To characterize the recovery of the active fraction of bacteria for each strain, the following empirical expression was proposed to fit the experimental data:
\begin{equation}
    \frac{\varphi(t_\text{ps})}{\varphi_0} = \varphi_\text{rec} - (\varphi_\text{rec} - \varphi_\text{ps}) e^{-t_\text{ps}/\tau_{\varphi}},
    \label{eq: model phi_A recovery}
\end{equation}
where $\varphi_0$ is the fraction of active bacteria in the control, non-sheared suspensions. In this expression, $\varphi(t_\text{ps} = 0)/\varphi_0 = \varphi_\text{ps}$, thus $\varphi_\text{ps}$ is the fraction of initially active bacteria that remain active immediately after the shearing process. Also, $\varphi(t_\text{ps} \to \infty)/\varphi_0 \to \varphi_\text{rec}$, thus $\varphi_\text{rec}$ is the fraction of bacteria that recover motility with time, conversely, $1 - \varphi_\text{rec}$, is the percentage of initially active bacteria in the suspensions whose BFMs were permanently damaged in the shearing process, making it impossible for them to recover their motility by regenerating their flagella.

The fitting parameters for each strain for the different experimental conditions are presented in Tab.~\ref{tab: models}. The fraction of remaining active bacteria, $\varphi_\text{ps}$ and the fraction of recovered bacteria, $\varphi_\text{rec}$ varied depending on the strain and shearing conditions, reflecting the effect of shear on the BFMs. Consistent with the observations in the previous section, we observed that, in both experiments, a larger fraction of WT bacteria remained active after the shearing process ($\varphi_\text{ps}$) in comparison with the $\Delta$\textit{lafA} mutant, presumably thanks to the protective effect of the lateral flagella, as proposed above. Similarly, the recovery percentage of active bacteria ($\varphi_\text{rec}$) was also larger for the WT strain. Again, this suggests that lateral flagella provided some degree of protection to the bacterial body, not only to the decrease in filament breakup but also against permanent damage to the flagella.

The regeneration times $\tau_\varphi$ should be interpreted as the characteristic time necessary for a bacterium whose flagellar filaments were completely, but reversibly cut, to regain motility, i.e., the time required for the BFMs of a cell to become functional again after a complete filament cut. The characteristic regeneration times of the active fraction obtained from the fitted curves were independent, within error bars, of the shearing conditions and shorter for the WT than for the $\Delta$\textit{lafA}. This suggests that the flagella regeneration rate of the WT strain is higher than that of $\Delta$\textit{lafA}. This can be verified by studying the recovery of the bacteria's mean swimming speed, which is proportional to flagellar filament length~\cite{52qin2012, 53higdon1979, 54higdon1979, lisevich2024}.

\begin{table*}[t]
	\begin{tabular}{>{\centering\arraybackslash}p{1.5cm}>{\centering\arraybackslash}p{1.5cm}>{\centering\arraybackslash}p{2cm}>{\centering\arraybackslash}p{1.5cm}>{\centering\arraybackslash}p{1.5cm}>{\centering\arraybackslash}p{2cm}>{\centering\arraybackslash}p{1.5cm}>{\centering\arraybackslash}p{1.5cm}>{\centering\arraybackslash}p{2cm}>{\centering\arraybackslash}p{2cm}}	
	Exp & $\varphi_0$ (\si{\percent}) & $\bar v_0$ (\si{\micro\meter/\second}) & $\varphi_\text{ps}$ (\si{\percent}) & $\varphi_\text{rec}$ (\si{\percent}) & $\tau_\varphi$ (\si{\minute}) & $\alpha_\text{sp}$ (\si{\percent}) & $\alpha_\text{lat}$ (\si{\percent}) & $\tau_\text{sp}$ (\si{\minute}) & $\tau_\text{lat}$ (\si{\minute}) \\
	\hline
	WT-d & \multirow{2}*{$22$} & \multirow{2}*{$22.09$} & 30 & 43 & $7 \pm 1$ & $6 \pm 2$ & $62 \pm 2$ & 289* &$46 \pm 4$\\
	WT--t& & & 15 & 24 & $6.2 \pm 0.1$ & $27 \pm 4$ & $19 \pm 3$ & 261* & $36 \pm 5$\\
	$\Delta$\textit{lafA}-d & \multirow{2}*{$19$} & \multirow{2}*{18.59} & 16 & 31 & $10 \pm 1$ & $66 \pm 1$ & -- & $289 \pm 8$ & --\\
	$\Delta$\textit{lafA}-t & & & 7 & 21 & $10 \pm 0.4 $ & $57 \pm 1$ & -- & $261 \pm 15$ & --\\
	\end{tabular}
    \caption{Fitting parameters for expressions Eqs.~\eqref{eq: model phi_A recovery} and \eqref{eq: model to fit WT v2}. Abbreviations: WT-d ($\Delta$\textit{lafA}-d): experiments with WT ($\Delta$\textit{lafA}) strain with constant shearing distance, WT-t ($\Delta$\textit{lafA}-t): experiments with WT ($\Delta$\textit{lafA}) strain with constant shearing time.\\
    *Obtained from the recovery curve of $\Delta$\textit{lafA}.\\
    }
    \label{tab: models}
\end{table*}

The mean swimming speed of the sheared suspensions of both strains also recovered with post-shear time (see Fig.~\ref{fig: recovery mot}(b)). Based on the hypothesis that the speed recovery occurs thanks to the elongation of the filaments of BFMs that were reversibly damaged, and assuming that those bacteria will eventually recover, on average, their original speed, then the expression that we propose is:
\begin{equation}
    \frac{\bar v(t_\text{ps})}{\bar v_0} = 1 - \alpha_\text{sp} e^{-t_\text{ps}/\tau_\text{sp}} - \alpha_\text{lat} e^{-t_\text{ps}/\tau_\text{lat}},
    \label{eq: model to fit WT v2}
\end{equation}
where $\bar v_0$ is the reference average swimming speed in the non-sheared suspensions. In Eq.~\eqref{eq: model to fit WT v2}, there are now two characteristic times: $\tau_\text{lat}$ associated with the regeneration of filaments in the lateral flagella and $\tau_\text{sp}$ associated with the regeneration of the filament of the subpolar flagellum. In contrast with $\tau_\varphi$, these times quantify the time required for full regeneration of the flagella, not only to regain functionality. Accordingly, $\alpha_\text{sp}$ and $\alpha_\text{lat}$ are the relative decrease in swimming speed associated with the cut of the subpolar and lateral flagella, respectively. Since the mutant strain $\Delta$\textit{lafA} only possesses the subpolar flagellum, it is possible to assume that the regeneration time is just $\tau_\text{sp}$ and that $\alpha_\text{lat} = 0$ in this case. For the WT strain, on the other hand, we assume that the subpolar flagellum recovers at the same rate that for the mutant strain, and only $\tau_\text{lat}, \alpha_\text{sp}$, and $\alpha_\text{lat}$ are left as adjustable parameters.

The fitting curves are shown as solid lines in Fig.~\ref{fig: recovery mot}(b), and the fitted parameters are presented in Tab.~\ref{tab: models}. The characteristic recovery time of each type of flagellar filament was comparable in the different shearing conditions. However, they were systematically longer when the cells were sheared for a constant distance and shorter time. The reason behind this is unclear, although at least for the WT, one possible reason is that the hypothesis of full-speed recovery is not fulfilled due to irreversible damage of some of the flagella. Comparing the two flagellar types, the filament regeneration time was notoriously shorter for the lateral flagella than for the subpolar one. This explains why the WT strain, which possesses both flagellar systems, recovers its motility faster than the $\Delta$\textit{lafA}, which possesses only a subpolar flagellum.

Finally, we note that for the WT in experiments with constant shearing distance, $\alpha_\text{sp}$ was significantly lower than $\alpha_\text{lat}$, meaning that the loss of swimming speed was mainly attributable to the damage in the lateral flagella in that case. In contrast, $\alpha_\text{sp}$ was larger than $\alpha_\text{lat}$ for constant shearing time. This would indicate more important damage in the subpolar flagellum due to the sustained shearing despite the protection of the lateral flagella. For the $\Delta$\textit{lafA} strain, the percentage of swimming speed loss, $\alpha_\text{sp}$, in both shearing conditions was larger than for the WT. The fact that it was smaller for constant shearing time than for constant shearing distance (\SI{57}{\percent} vs. \SI{66}{\percent}) could be due to the smaller shear used in the experiments with constant shearing time. This is consistent with the observations in the previous section, where the loss in swimming speed and active fraction for the $\Delta$\textit{lafA} was independent of the shearing conditions and only depended on the value of shear.

\section{Discussion}

The picture that emerges from our experimental results is one where soil bacteria, usually exposed to shear flows that partially cut their flagellar filaments, possess a recovery mechanism to maintain the efficiency of swimming motility. Of the two flagellar systems that \textit{B. diazoefficiens} possess, the subpolar flagellum is thicker and appears to be the fundamental actuator for swimming motility, as suggested by the poor swimming behavior of the mutants lacking it~\cite{19quelas2016, MonteiroPreprint}. At the same time, our observations with the $\Delta$\textit{lafA} mutant suggest that this flagellum is highly sensitive to shear, with this strain being rapidly affected by shear even for very short durations of merely \SI{0.06}{\second} for sufficiently large shear rates, displaying a dramatic drop of actively swimming bacteria and a lower recovery fraction. In comparison, motility is better conserved in the WT strain when exposed to shear, suggesting that, contrary to our original expectations, the thin and long lateral flagella can better stand sustained shear than the thicker subpolar flagellum.

On the other hand, our measurements of motility recovery demonstrate a much shorter recovery time for the lateral flagella than for the subpolar one. This is consistent with the lower molecular weight of lateral flagellins (\SI{34}{\kilo\dalton} for lateral flagellins \textit{vs.} \SI{68}{\kilo\dalton} for the subpolar flagellins~\cite{19quelas2016}), making their diffusion much faster. This observation can also be related to the fact that the metabolic rate of the WT is approximately three times higher than the metabolic rate of the $\Delta$\textit{lafA} mutant since the WT has to continuously regenerate 3 to 5 lateral flagella more than the $\Delta$\textit{lafA} mutant, which only regenerates the subpolar flagellum~\cite{Cogo2018}. These filaments may undergo continuous breakup, countered by filament growth, even in the presence of moderate shear rates, such as the continuous shaking during incubation, and this difference in metabolic rate reflects the continuous synthesis of lateral flagellins.

It has been observed that the subpolar flagellum acts as a mechanosensor that regulates the induction of lateral flagella, i.e., destabilization of the subpolar flagellum induces a higher synthesis of lateral flagella~\cite{18mengucci2020}. This stimulus is unidirectional, and the breakup of lateral flagella does not influence the expression of the subpolar flagellum. Taken together, these observations suggest that lateral flagella act as a protection mechanism for the subpolar flagellum.

Our work provides a better understanding of the function of both flagellar systems for the swimming motility of \textit{Bradyrhizobium diazoefficiens}. Many of the conclusions, however, are expected to qualitatively apply to other bacterial species that rely on BFMs for their motility. Indeed, bacteria can be exposed to shear flows in a multitude of situations, either in their natural habitats due to rainfall, artificial irrigation systems, marine waves, physiological flows in their hosts, etc., or in the laboratory due to agitation during incubation and when studied in artificial microfabricated devices~\cite{Altshuler2012, Scheidweiler2020, MonteiroPreprint}. The effect could be even enhanced in complex environments such as non-Newtonian fluids or crowded systems~\cite{Bechinger2016}. This can have important consequences for biological and active matter systems where the effects of time-dependent motility have been little studied, for example, in the efficiency of ratchet-like geometries for bacterial sorting~\cite{Galajda2007, Wan2008, Berdakin2013a, Berdakin2013b,OlsonReichhardt2017} and for the extraction of useful work from bacterial suspensions~\cite{DiLeonardo2010, Sokolov2010, Vincenti2019}.

In conclusion, our work studied the interaction between bacterial flagella and fluid flows in long and thin microchannels and demonstrated that simple measurements of bacteria motility can be helpful to understand better the two flagellar systems of \textit{B. diazoefficiens}, a bacterium widely used as a biofertilizer. In this way, applied physics of fluids and microfluidics, both theoretical and experimental, can give again, as in Refs.~\onlinecite{cortezMaster2014, 19quelas2016, montagnaMaster2018, gutierrezMaster2023, MonteiroPreprint},  important fundamental insights in other fields such as microbiology. The inclusion of other factors, such as an increased medium viscosity or even non-Newtonian effects that better represent complex environments present in real soil, can be easily included in this kind of setup.

\section*{Acknowledgments}
This research was funded by ANID - Millennium Science Initiative Program NCN19\_170. VIM acknowledges support from grants: SeCyT-UNC: 33620230100298CB; FONCyT: PICT-2020-SERIEA-02931; CONICET: PIP-2023-11220220100509CO. Fabrication of microfluidic devices was possible thanks to ANID Fondequip grants Nos. EQM140055 and EQM180009. JPCM acknowledges funding from ANID Beca de Magíster Nacional No. 22221639. MPM acknowledges the postdoctoral Fondecyt Grant No. 3190637.

\section*{Author declarations}

\subsection*{Conflict of interest }
The authors have no conflicts to disclose.

\subsection*{Author contributions}
\textbf{Juan Pablo Carrillo-Mora:} Conceptualization (equal); Methodology (lead): Investigation (lead); Formal analysis (lead); Writing -- original draft (lead); Writing -- review \& editing (equal). \textbf{Moniellen Pires Monteiro:} Conceptualization (equal); Supervision (supporting). \textbf{Aníbal R. Lodeiro:} Conceptualization (equal); Writing -- review \& editing (equal). \textbf{V. I. Marconi:} Conceptualization (equal); Software (lead); Writing -- review \& editing (equal). \textbf{María Luisa Cordero:} Conceptualization (equal); Supervision (lead); Writing -- original draft (supporting); Writing -- review \& editing (equal).

\section*{Data accessibility}
The data that support the findings of this study are available from the corresponding author upon reasonable request.


%

\end{document}